\documentstyle{mn}

\newif\ifAMStwofonts

\ifoldfss
  \ifCUPmtlplainloaded \else
    \NewTextAlphabet{textbfit} {cmbxti10} {}
    \NewTextAlphabet{textbfss} {cmssbx10} {}
    \NewMathAlphabet{mathbfit} {cmbxti10} {} 
    \NewMathAlphabet{mathbfss} {cmssbx10} {} 
  \fi
  \ifAMStwofonts
    \ifCUPmtlplainloaded \else
      \NewSymbolFont{upmath} {eurm10}
      \NewSymbolFont{AMSa} {msam10}
      \NewMathSymbol{\upi}     {0}{upmath}{19}
      \NewMathSymbol{\umu}     {0}{upmath}{16}
      \NewMathSymbol{\upartial}{0}{upmath}{40}
      \NewMathSymbol{\leqslant}{3}{AMSa}{36}
      \NewMathSymbol{\geqslant}{3}{AMSa}{3E}

       \let\le=\leqslant
       
    \fi
  \fi
\fi 

\ifnfssone
  \newmathalphabet{\mathit}
  \addtoversion{normal}{\mathit}{cmr}{m}{it}
  \addtoversion{bold}{\mathit}{cmr}{bx}{it}
  \newmathalphabet{\mathbfit} 
  \addtoversion{normal}{\mathbfit}{cmr}{bx}{it}
  \addtoversion{bold}{\mathbfit}{cmr}{bx}{it}
  \newmathalphabet{\mathbfss} 
  \addtoversion{normal}{\mathbfss}{cmss}{bx}{n}
  \addtoversion{bold}{\mathbfss}{cmss}{bx}{n}
  \ifAMStwofonts
    \ifCUPmtlplainloaded \else
      \UseAMStwoboldmath
      \makeatletter
      \new@mathgroup\upmath@group
      \define@mathgroup\mv@normal\upmath@group{eur}{m}{n}
      \define@mathgroup\mv@bold\upmath@group{eur}{b}{n}
      \edef\UPM{\hexnumber\upmath@group}
      \new@mathgroup\amsa@group
      \define@mathgroup\mv@normal\amsa@group{msa}{m}{n}
      \define@mathgroup\mv@bold\amsa@group{msa}{m}{n}
      \edef\AMSa{\hexnumber\amsa@group}
      \makeatother
      \mathchardef\upi="0\UPM19
      \mathchardef\umu="0\UPM16
      \mathchardef\upartial="0\UPM40
      \mathchardef\leqslant="3\AMSa36
      \mathchardef\geqslant="3\AMSa3E

       \let\le=\leqslant

    \fi
  \fi
\fi 

\ifnfsstwo
  \DeclareMathAlphabet{\mathbfit}{OT1}{cmr}{bx}{it}
  \SetMathAlphabet\mathbfit{bold}{OT1}{cmr}{bx}{it}
  \DeclareMathAlphabet{\mathbfss}{OT1}{cmss}{bx}{n}
  \SetMathAlphabet\mathbfss{bold}{OT1}{cmss}{bx}{n}
  \ifAMStwofonts
    \ifCUPmtlplainloaded \else
      \DeclareSymbolFont{UPM}{U}{eur}{m}{n}
      \SetSymbolFont{UPM}{bold}{U}{eur}{b}{n}
      \DeclareSymbolFont{AMSa}{U}{msa}{m}{n}
      \DeclareMathSymbol{\upi}{0}{UPM}{"19}
      \DeclareMathSymbol{\umu}{0}{UPM}{"16}
      \DeclareMathSymbol{\upartial}{0}{UPM}{"40}
      \DeclareMathSymbol{\leqslant}{3}{AMSa}{"36}
      \DeclareMathSymbol{\geqslant}{3}{AMSa}{"3E}

       \let\le=\leqslant

    \fi
  \fi
\fi 

\ifCUPmtlplainloaded \else
  \ifAMStwofonts \else 
    \def\upi{\pi}
    \def\umu{\mu}
    \def\upartial{\partial}
  \fi
\fi

\title{New clues to the evolution of dwarf early--type galaxies}
\author[D. Pierini]
       {D. Pierini \\
        Department of Physics and Astronomy, The University of Toledo,
        Toledo, OH 43606}
\date{Accepted ...
      Received ...;
      in original form ...}

\pagerange{\pageref{firstpage}--\pageref{lastpage}}
\pubyear{...}

\begin{document}

\maketitle

\label{firstpage}

\begin{abstract}
Surface photometry of 18 Virgo cluster dwarf elliptical (dE)
and dwarf lenticular (dS0) galaxies, made by Gavazzi et al. (2001)
in the H-band ($\rm 1.65~\mu m$) and in the B-band ($\rm 0.44~\mu m$),
shows that the ratio of the effective radii of these stellar systems
in the B- and H-band, $r_{e B}/r_{e H}$, ranges between 0.7 and 2.2.
In particular, dwarf ellipticals and lenticulars with
a red total color index B$-$H (i.e. with $3.2 <$ B$-$H $< 4$)
have equal effective radii in these two pass-bands.
By contrast, blue (i.e. with $2.5 <$ B$-$H $< 3.1$) dEs and dS0s have
B-band effective radii about 50\% longer than the H-band ones, on average.
Consistently, strong negative gradients in B$-$H along the galactocentric
radius are found to be associated with blue total colors.
This trend is not found in a sample of 29 giant E and S0 galaxies
of the Coma cluster with analogous data available in the literature.
These early-type giants span a broad range in $r_{e B}/r_{e H}$ (0.2--2.2),
with a mean $r_{e B}/r_{e H} \sim 1.1$, but a narrow range in (red) color
($3.3 <$ B$-$H $< 4.2$).
In these stellar systems, color gradients are usually interpreted
as due either to age/metallicity gradients along the radial coordinate
or to dust attenuation, whatever the total color of the system is.
Assuming each of these three distinct interpretations of the origin
of color gradients, we discuss the origin of the association
of strong negative color gradients with blue colors found
in the early-type dwarfs under study, in relation with current scenarios
of formation and evolution of dE and dS0 galaxies.
\end{abstract}


\section{Introduction}

Dwarf elliptical (dE) and dwarf lenticular (dS0) galaxies
are the most common galaxies in the local universe, as discovered by
the earliest studies of nearby clusters, such as Virgo
(Binggeli, Sandage \& Tammann 1985; Sandage, Binggeli \& Tammann 1985)
and Fornax (Ferguson \& Sandage 1988; Ferguson 1989).
Nonetheless, their origin (e.g. White \& Frenk 1991; Ostriker 1993;
Babul \& Rees 1992; Koo et al. 1995, 1997; Mao \& Mo 1998;
Moore, Lake \& Katz 1998; Moore et al. 1999; Lin \& Faber 1983)
is still matter of uncertainty
(e.g. Conselice, Gallagher \& Wyse 2001; Drinkwater et al. 2001).
An exponential light distribution seems to be characteristic
of early-type dwarfs, though not ubiquitous, both in the optical
(Faber \& Lin 1983) and in the near-IR (James 1991, 1994).
However, it is still matter of debate whether this stellar disk-component 
is rotationally flattened or not (e.g. Rix, Carollo \& Freeman 1999;
Geha, Guhathakurta \& van der Marel 2001).

As a latest result of their extensive observational campaign centered
on H-band ($\rm 1.65~\mu m$) imaging of about 1200 nearby giant and dwarf
early- and late-type galaxies, Gavazzi and collaborators
(Gavazzi et al. 2000, 2001) have found that the decomposition
of the H-band surface brightness profile of a galaxy
is a strong function of its total H-band luminosity,
whatever the Hubble classification of the galaxy is.
In fact, the fraction of exponential-disk law (Freeman 1970) decompositions
decreases with increasing luminosity of the galaxy, while the fraction
of bulge$+$disk decompositions increases with luminosity.
In particular, pure de Vaucouleurs-law (de Vaucouleurs 1948) profiles
are absent at near-IR luminosities lower than $\rm 10^{10}~L_{\odot}$.
In this regime, Gavazzi et al. (2001) find that dwarf-elliptical peculiar
(pec) galaxies have structural parameters indistinguishable from
those of late-type dwarfs and, therefore, propose that dE pec galaxies
represent the missing link between dEs and dwarf irregulars (dIs).

Though the presence of an exponential-disk component in dE, dS0
and dI galaxies, in general, may point to similarities between
the formation and evolution processes of these three types of dwarf galaxies,
only the late-type galaxies are definitely held to be HI-rich
(e.g. Boselli et al. 2001).
In fact, there is recent evidence that some dE galaxies are associated
with atomic gas, but HI gas and stars may not always be parts
of the same dynamical system (e.g. Young \& Lo 1997).
Atomic gas plays a fundamental role in the formation and evolution
of exponential-disks (e.g. Dalcanton, Spergel \& Summers 1997;
Boissier \& Prantzos 2000; Ferguson \& Clarke 2001).
As reviewed by the latter authors, the origin of stellar exponential-disks
finds a natural explanation in viscosity-driven radial flows,
under the assumption that the star formation and viscous timescales
are comparable.
Assuming this simultaneity, the exponential-disk scale-length
is found either to increase with time, if the cosmologically-motivated
gaseous infall is concurrent with the previous two processes,
or to stay constant if the bulk of the disk is assembled before
star formation and viscosity act in a significant manner
(Ferguson \& Clarke 2001).
In the former case, successive stellar populations trace
the settling of the newly infalling gas at different disk ages.
As a straightforward consequence, a radial color gradient is produced
and the exponential-disk scale-lengths are expected to be larger
in the optical pass-bands than in the near-IR ones.

However, not only disk-galaxies but also disk-less galaxies 
may show a wavelength-dependence of their characteristic scale-length
(e.g. the effective radius, i.e. the radius which contains 50\%
of the total luminosity), simply as a result of the well-known effects
of age and metallicity on broad-band colors of stellar populations
(e.g. Bruzual \& Charlot 1993; Worthey 1994). 
In fact, color gradients along the radial coordinate of a galaxy may be
due either to a gradient in age (with fixed metallicity) or to a gradient
in metallicity (with fixed age) of its stellar populations
(e.g. Peletier, Valentijn \& Jameson 1990; Tamura et al. 2000;
Saglia et al. 2000; de Jong 1996), whatever the causes of these gradients are.
Furthermore, dust attenuation toward the galaxy center is able to reproduce
differences in the observed radial surface brightness distribution
at two wavelengths (e.g. in the optical and near-IR), due to the difference
in optical depth at these two wavelengths for a given dust column density,
even under the hypothesis that a simple stellar population is present,
and without introducing a significant change in the observed total color
(Witt, Thronson \& Capuano 1992).

In order to investigate the presence of color gradients and their origin
(if any) in early-type dwarfs, here we study the relationships (if any)
between different photometric parameters (effective radius $r_e$,
bulge-to-total luminosity ratio $B/T$, total optical--near-IR color index
B$-$H and B$-$H color profile) of 18 Virgo cluster dE and dS0 galaxies,
imaged by Gavazzi et al. (2001) in the B-band ($\rm 0.44~\mu m$) and H-band.
This sample comprises objects one to a few magnitudes brighter
than dwarf elliptical and spheroidal galaxies seen in the Local Group
(e.g. Mateo 1998), which form a different population of early-type dwarfs
(e.g. Conselice, Gallagher \& Wyse 2001 and references therein),
as witnessed also by their different photometric parameters.

\section{The sample}

B- ($\rm 0.44~\mu m$), V- ($\rm 0.55~\mu m$) and H-band ($\rm 1.65~\mu m$)
surface photometry of 17 dwarf elliptical and lenticular galaxies
plus 1 low-mass E pec/S0 galaxy was obtained by Gavazzi et al.
(2001 -- hereafter referred to as G01).
These objects, with photographic magnitude $\rm m_p \le 16.0$,
were selected from the Virgo Cluster Catalogue (VCC) of
Binggeli, Sandage \& Tammann (1985).
With the exception of VCC\,1078, they are all certain members of the cluster,
where membership was assigned by Binggeli, Sandage \& Tammann
and Binggeli, Popescu \& Tammann (1993).
Hereafter we refer to these 18 dwarf/low-mass early-type galaxies
as the VCC early-type dwarf sample.
\begin{figure}
 \vspace{400pt}
 \caption{Projected spatial distribution of the 18 VCC early-type dwarfs
  listed in Tab. 1 on to the sky-region occupied by the Virgo cluster.
  We indicate the position of the maximum projected galaxy density
  of cluster A (containing M\,87), of M\,87 and of the centre of cluster B
  (containing M\,49) as given by Binggeli, Tammann \& Sandage (1987),
  as well as the sky-regions defined as the Southern Extension,
  M and W Clouds (see Binggeli, Tammann \& Sandage).
  Furthermore, we draw the circle of 4 degrees radius centred on the centre
  of Cluster A, the circle of 2 degrees radius centred on M\,87
  and the circle of 1.5 degrees radius centred on the centre of Cluster B.}
\end{figure}

Fig. 1 gives a pictorial view of the projected spatial distribution
of these galaxies on to the sky-region occupied by the Virgo cluster.
It shows that the dwarfs under study lay either within a projected
radial distance of 2 degrees from M\,87 (associated with cluster A),
with the exception of VCC\,608 (laying in the corona between
the previous region and a circle of 4 degrees radius centred
on the maximum projected galaxy density of cluster A),
or within a projected radial distance of 1.5 degrees from the centre
of cluster B (see Binggeli, Tammann \& Sandage 1987 for the definitions
of clusters A and B).
\begin{table*}
 \centering
 \begin{minipage}{130mm}
  \caption{Catalogue parameters of the Virgo cluster early-type dwarfs.}
  \begin{tabular}{@{}rlcclccr@{}}
   VCC &~NGC/IC & RA(1950.0) & Dec(1950.0) &~~~~~~~~Hubble & log~D & log~R & v~~~~~~\\
   Den. &~~~~Den. & hh~~mm~~ss & $^o$~~$^{\prime}$~~$^{\prime \prime}$ &~~~~~~~~~~type & & & $\rm km~s^{-1}$\\[10pt]
 608 & NGC\,4323 & 12~~20~~29.70 & 16~~10~~58.0 & dE4,N & 1.16 & 0.30 & 1803~~~ \\
 745 & NGC\,4366 & 12~~22~~14.40 & 07~~37~~48.0 & dE6,N & 1.03 & 0.38 & 1286~~~ \\
 786 & IC\,3305 & 12~~22~~43.80 & 12~~07~~54.0 & dE7,N & 1.32 & 0.58 & 2388~~~ \\
 951 & IC\,3358 & 12~~24~~22.20 & 11~~56~~42.0 & dE2,N pec/dS0(2),N & 1.28 & 0.18 & 2066~~~ \\
 965 & IC\,3363 & 12~~24~~31.20 & 12~~50~~06.0 & dE7,N & 1.26 & 0.52 & 790~~~ \\
1036 & NGC\,4436 & 12~~25~~10.20 & 12~~35~~30.0 & dE6/dS0(6),N & 1.33 & 0.38 & 1163~~~ \\
1073 & IC\,794 & 12~~25~~36.50 & 12~~22~~11.0 & dE3,N & 1.26 & 0.12 & 1899~~~ \\
1078 & & 12~~25~~39.00 & 10~~02~~24.0 & dE5 pec? & 0.86 & 0.30 & \\
1122 & IC\,3393 & 12~~26~~09.60 & 13~~11~~30.0 & dE7,N & 1.30 & 0.64 & 436~~~ \\
1173 & & 12~~26~~43.00 & 13~~15~~16.0 & dE5,N & 0.80 & 0.29 & 2468~~~ \\
1254 & NGC\,4472 & 12~~27~~32.80 & 08~~21~~03.0 & dE0,N & 0.82 & 0.00 & 1350~~~ \\
1308 & IC\,3437 & 12~~28~~14.40 & 11~~37~~00.0 & dE6,N & 1.10 & 0.45 & 1721~~~ \\
1348 & IC\,3443 & 12~~28~~43.90 & 12~~36~~28.0 & dE0,N pec & 0.77 & 0.00 & 1679~~~ \\
1386 & IC\,3457 & 12~~29~~19.20 & 12~~56~~00.0 & dE3,N & 1.17 & 0.22 & 1426~~~ \\
1453 & IC\,3478 & 12~~30~~12.80 & 14~~28~~19.0 & dE2,N & 1.21 & 0.11 & 1949~~~ \\
1491 & IC\,3486 & 12~~30~~42.40 & 13~~08~~00.0 & dE2,N & 0.86 & 0.12 & 903~~~ \\
1499 & IC\,3492 & 12~~30~~48.20 & 13~~07~~44.0 & E3 pec/S0 & 0.80 & 0.15 & \\
1684 & IC\,3578 & 12~~34~~07.80 & 11~~22~~42.0 & dS0(8) & 1.10 & 0.54 & \\
  \end{tabular}
 \end{minipage}
\end{table*}
The 3D structure of the Virgo cluster is complex (e.g. de Vaucouleurs 1961).
According to Gavazzi et al. (1999), the distance modulus $\rm \mu_0$
of cluster A is $\rm 30.84 \pm 0.06$, while, for cluster B, dominated
by M\,49, $\rm \mu_0 = 31.84 \pm 0.10$.
As a consequence, the galaxies associated with cluster B are 60\%
farther from us than those associated with cluster A.
For the purposes of the present analysis, we assume that all
the 18 VCC galaxies under study lay at a distance of 17.0 Mpc to us,
in agreement with G01.

The VCC early-type dwarf sample contains 13 dE,N galaxies,
1 dE,N/dS0,N galaxy and 1 dE/dS0,N galaxy, where N stands for ``nucleated''.
Such a high fraction of nucleated dwarf ellipticals reflects the fact that
dE,N galaxies are more luminous than non-nucleated
dwarf ellipticals and, therefore, their fraction increases
in a magnitude-limited sample such as the G01 one.
It is also straigthforward to understand that dE,Ns are the easiest
to detect at both optical and near-IR pass-bands among dEs.
Though there are different possibilities of producing nucleation, 
Conselice, Gallagher \& Wyse (2001) conclude that dEs and dE,Ns
within 6 degrees from the centre of Virgo cluster A
and with heliocentric radial velocity $\rm v < 2400~km~s^{-1}$
have similar origins, since the velocity characteristics
of these two populations are not significantly different.
In particular, these authors list heliocentric radial velocities for 15
out of the 18 galaxies of the VCC early-type dwarf sample.
Relying on their result, we assume that all the 18 galaxies under study
have the same origin.

In Tab. 1, we list the VCC catalogue properties and the heliocentric
radial velocities (Conselice, Gallagher \& Wyse 2001) of
the individual galaxies relevant to this study, as follows: \newline
Col. 1: the galaxy denomination; \newline
Col. 2: alternate (NGC/IC) galaxy denomination; \newline
Col. 3,4: celestial coordinates (RA and Dec, respectively)
at the Equinox B1950; \newline
Col. 5: the morphological classification; \newline
Col. 6: decimal logarithm of the major-axis diameter D measured
on the du Pont plates at the faintest detectable isophote
(in units of $\rm 0.1^{\prime}$); \newline
Col. 7: estimated major-to-minor axis ratio R, measured analogously to D;
\newline
Col. 8: the heliocentric radial velocity.
\begin{table*}
 \centering
 \begin{minipage}{160mm}
  \caption{Photometric parameters of the Virgo cluster early-type dwarfs.}
  \begin{tabular}{@{}rrrrccrrrcccc@{}}
   VCC & $\rm r_{ebf H}$ & $\rm r_{edf H}$ &
   $\rm r_{e H}~~~~$ & cl. & $\rm B/T_H$ & $\rm r_{ebf B}$ & $\rm r_{edf B}$ &
   $\rm r_{e B}$~~~~ & cl. & $\rm B/T_B$ & $\rm H$~ & $\rm B$~ \\
   Den. & $\rm ^{\prime \prime}~~$ & $\rm ^{\prime \prime}$~~~&
   $\rm ^{\prime \prime}$~~~~~~& & & $\rm ^{\prime \prime}$~~~&
   $\rm ^{\prime \prime}$~~~& $\rm ^{\prime \prime}$~~~~~~& & & mag~ & mag~ \\[10pt]
 608 & & 17.26 & 17.7$\pm$1.1 & 3 & 0.00 & & 16.47 & 15.7$\pm$1.2 & 3 & 0.00 & 11.87$\pm$0.06 & 15.29$\pm$0.07\\
 745 & 8.99 & & 8.8$\pm$1.1 & 1 & 1.00 & 6.53 & 24.36 & 16.3$\pm$1.4 & 2 & 0.29 & 11.99$\pm$0.06 & 14.95$\pm$0.06\\
 786 & & 16.20 & 15.6$\pm$1.4 & 3 & 0.00 & 10.25 & 32.13 & 27.8$\pm$3.0 & 2 & 0.29 & 12.21$\pm$0.06 & 14.84$\pm$0.11\\
 951 & 3.99 & 18.33 & 17.2$\pm$1.3 & 2 & 0.04 & 6.63 & 25.85 & 21.1$\pm$1.6 & 2 & 0.26 & 11.60$\pm$0.06 & 14.46$\pm$0.07\\
 965 & 1.83 & 29.34 & 30.1$\pm$7.9 & 2 & 0.01 & 0.92 & 21.67 & 23.0$\pm$1.4 & 2 & 0.01 & 11.66$\pm$0.31 & 15.54$\pm$0.06\\
1036 & 7.98 & 30.02 & 18.5$\pm$1.5 & 2 & 0.36 & 7.82 & 31.01 & 21.6$\pm$1.6 & 2 & 0.28 & 10.65$\pm$0.06 & 14.06$\pm$0.06\\
1073 & 2.77 & 12.30 & 10.8$\pm$1.1 & 2 & 0.11 & 5.14 & 26.09 & 24.6$\pm$1.5 & 2 & 0.20 & 11.41$\pm$0.06 & 14.48$\pm$0.06\\
1078 & & 11.44 & 11.7$\pm$1.1 & 3 & 0.00 & & 12.73 & 12.2$\pm$1.1 & 3 & 0.00 & 13.37$\pm$0.06 & 16.01$\pm$0.06\\
1122 & 12.37 & & 11.8$\pm$1.0 & 1 & 1.00 & 9.05 & 28.88 & 16.6$\pm$1.4 & 2 & 0.43 & 11.99$\pm$0.06 & 15.00$\pm$0.06\\
1173 & & 11.77 & 12.1$\pm$1.1 & 3 & 0.00 & & 10.61 & 11.3$\pm$1.1 & 3 & 0.00 & 13.20$\pm$0.06 & 16.43$\pm$0.06\\
1254 & 0.93 & 13.02 & 12.2$\pm$1.4 & 2 & 0.05 & 1.44 & 13.02 & 12.3$\pm$1.2 & 2 & 0.22 & 12.07$\pm$0.06 & 15.73$\pm$0.06\\
1308 & 7.84 & & 7.5$\pm$1.0 & 1 & 1.00 & 6.87 & 19.25 & 11.1$\pm$1.2 & 2 & 0.44 & 12.73$\pm$0.06 & 15.75$\pm$0.07\\
1348 & 1.19 & 8.26 & 7.7$\pm$1.0 & 2 & 0.06 & 1.79 & 9.58 & 8.4$\pm$1.1 & 2 & 0.09 & 12.42$\pm$0.06 & 16.01$\pm$0.07\\
1386 & 19.79 & & 20.9$\pm$1.2 & 1 & 1.00 & 16.57 & 52.87 & 34.7$\pm$4.0 & 2 & 0.35 & 12.07$\pm$0.06 & 14.82$\pm$0.10\\
1453 & 3.07 & 10.86 & 10.1$\pm$1.1 & 2 & 0.10 & 4.79 & 20.40 & 19.1$\pm$1.4 & 2 & 0.18 & 11.58$\pm$0.07 & 14.46$\pm$0.07\\
1491 & 7.23 & 36.41 & 15.1$\pm$5.1 & 2 & 0.46 & 7.78 & 28.10 & 13.3$\pm$1.3 & 2 & 0.41 & 11.64$\pm$0.20 & 15.26$\pm$0.07\\
1499 & 7.21 & & 7.7$\pm$1.0 & 1 & 1.00 & 4.92 & 12.67 & 7.4$\pm$1.1 & 2 & 0.64 & 12.57$\pm$0.07 & 15.18$\pm$0.07\\
1684 & 2.08 & 19.76 & 20.6$\pm$1.1 & 2 & 0.01 & 1.95 & 17.39 & 20.3$\pm$1.5 & 2 & 0.03 & 12.64$\pm$0.06 & 15.46$\pm$0.06\\
\end{tabular}
\end{minipage}
\end{table*}

We refer the reader to G01 for details concerning observations,
data reduction, photometric calibration and image reduction procedures
of the 18 VCC galaxies under study.
For the present analysis it is important to say that these authors
derived azimuthally averaged surface brightness profiles, that were fitted
using either a de Vaucouleurs $r^{1/4}$ law, an exponential-law,
a bulge$+$disk model or an exponential/de Vaucouleurs truncated model.
In particular, the total magnitude of each galaxy was obtained
by adding to the flux measured within the outermost significant isophote
the flux extrapolated to infinity along the model that fitted
the outer parts of the galaxy.
The effective radius $r_e$ and the effective surface brightness
$\mu_e$ (i.e. the mean surface brightness within $r_e$)
of each galaxy were computed in two ways (Gavazzi et al. 2000).
The ``fitted'' values of the individual components were derived from
the individual fitted profiles, extrapolated to zero and to infinity.
By contrast, the ``empirical'' values of the effective radius
and surface brightness of the system were obtained
locating the half-light point along the observed light profile,
where the total amount of light is given by the total magnitude
described above, and corrected for seeing according to Saglia, Bender
\& Dressler (1993).
The ``empirical'' values of $r_e$ determined by G01 are used here. 
Finally, the G01 determined the bulge-to-total luminosity ratio
of individual galaxies as a result of the fitting procedure.

Tab. 2 contains the H- and B-band photometric properties
of individual galaxies listed in G01 and relevant to this study,
as follows: \newline
Col. 1: the galaxy denomination in the VCC; \newline
Col. 2: the H-band bulge effective radius $r_{ebf H}$; \newline
Col. 3: the H-band disk effective radius $r_{edf H}$; \newline
Col. 4: the H-band effective radius $r_{e H}$ and its error; \newline
Col. 5: the classification according to the H-band profile decomposition
($1 =$ de Vaucouleurs-law, $2 =$ bulge$+$disk system, $3 =$ exponential-law,
$4 =$ exponential/de Vaucouleurs truncated model); \newline
Col. 6: the H-band bulge-to-total luminosity ratio $B/T_H$
(0.00 and 1.00 identify bulge-less and disk-less systems, respectively);
\newline
Col. 7: the B-band bulge effective radius $r_{ebf B}$; \newline
Col. 8: the B-band disk effective radius $r_{edf B}$; \newline
Col. 9: the B-band effective radius $r_{e B}$ and its error; \newline
Col. 10: the classification according to the B-band profile decomposition
(see Col. 5); \newline
Col. 11: the B-band bulge-to-total luminosity ratio $B/T_B$ (see Col. 6);
\newline
Col. 12: the observed total H-band magnitude H and its error; \newline
Col. 13: the observed total B-band magnitude B and its error.

We note that 5 objects (i.e.: VCC\,745, 1122, 1308, 1386 and 1499)
change classification from a de Vaucouleurs- to a bulge$+$disk system,
and 1 object (VCC\,786) changes its classification from an exponential-disk
to a bulge$+$disk system, from the near-IR to the optical case.
The B-band surface brightness profile of VCC\,786 is decomposed
into relatively extended and equally faint bulge- and disk-components
(cf. G01), so that its profile fitting may be questionable
but not its values of B and $r_{e B}$.
In the case of VCC\,1499, the H-band surface brightness profile extends up
to 70\% of the radial extension of its B-band photometry,
so that the photometric parameters of this object are reliable.
By contrast, for VCC\,745, 1122, 1308 and 1386, the H-band surface brightness
profiles extend only up to 30--50\% of the radial extension
of the B-band ones.
For these 4 galaxies, it may be questioned that a faint, extended
exponential-disk might not have been detected against the near-IR background
(typically of 20.5 H-mag $\rm arcsec^{-2}$ in the images of these galaxies).
If an exponential-disk component was indeed missed/misidentified
by the near-IR observations, a best-fit de Vaucouleurs-law would
underestimate the surface intensity in the outer regions of these systems.
The visual inspection of the radial photometric profiles
of VCC\,745, 1122, 1308 and 1386 displayed in G01 suggests that
the previous systematic effect, if any, is not strong.
We estimate that the H-band total magnitudes of these 4 galaxies,
listed in Col. 11 of Tab. 2, may be under-luminous by $\rm \sim 0.2~mag$,
while their values of $r_{e H}$ may be underestimated by $\sim$10\%.
We anticipate that these potential systematic uncertainties do not bias
our results (cf. Sect. 3).

In addition to these 18 VCC early-type dwarfs, we consider, as a comparison,
a sample of 29 giant elliptical and lenticular galaxies of the Coma cluster,
optically selected from the Catalogue of Galaxies and Clusters of Galaxies
(CGCG) of Zwicky et al. (1961--1968) and observed in the H-band by
Scodeggio et al. (1998 -- hereafter referred to as S98).
We adopt a distance from the Coma cluster to us of 96.0 Mpc from S98.
The H-band data reduction and analysis of S98 is the same of G01,
except that all the radial surface brightness profiles of the S98 galaxies
were fitted with a de Vaucouleurs-law.
These 29 CGCG early-type giants were observed in the B-band
by different authors, listed in S98, adopting either photoelectric aperture
photometry or CCD surface photometry, but their B-band photometric parameters
were derived by Saglia, Bender \& Dressler (1993) by fitting
a de Vaucouleurs-law to the radial photometric profile derived from 2D data.
Therefore, for these giants, only fitted effective radii are available.
Both observed total H- and B-band magnitudes of these 29 galaxies
come from G. Gavazzi (2001, private communication).
These H-band total magnitudes were derived by S98 by adopting
the same procedure of G01.
By contrast, these B-band total magnitudes were either taken from
the literature, if derived from surface photometry, or derived by
Gavazzi \& Boselli (1996) by applying the ``growth curves'' technique
to the available photoelectric aperture measurements.
Both H- and B-band total observed magnitudes have a typical uncertainty
of 0.1 magnitude.
\begin{table*}
 \centering
 \begin{minipage}{150mm}
  \caption{Characteristic parameters of 29 giant E and S0 galaxies of the Coma cluster.}
  \begin{tabular}{@{}lllrccrcccc@{}}
   CGCG & NGC/IC & Hubble & $\rm r_{e H}~~~~$ & cl. & $\rm B/T_H$ &
   $\rm r_{e B}$~~~~ & cl. & $\rm B/T_B$ & $\rm H$~ & $\rm B$~ \\
   ~~Den. &~~~Den. & ~~type & $\rm ^{\prime \prime}$~~~~~~& & &
   $\rm ^{\prime \prime}$~~~~~~& & & mag~ & mag~ \\[10pt]
160017 & NGC\,4807 & S0 pec &  7.53$\pm$0.99 & 1 & 1.00 &  8.95$\pm$2.70 & 1 & 1.00 & 10.69$\pm$0.1 & 14.56$\pm$0.1\\
160021 & NGC\,4816 & S0     & 14.63$\pm$0.72 & 1 & 1.00 & 20.99$\pm$2.70 & 1 & 1.00 & 10.06$\pm$0.1 & 13.74$\pm$0.1\\
160028 & NGC\,4827 & S0     & 10.28$\pm$0.70 & 1 & 1.00 & 13.87$\pm$2.41 & 1 & 1.00 & 10.37$\pm$0.1 & 14.10$\pm$0.1\\
160042 & NGC\,4840 & S0     &  6.32$\pm$0.48 & 1 & 1.00 &  6.79$\pm$2.41 & 1 & 1.00 & 10.84$\pm$0.1 & 14.73$\pm$0.1\\
160063 & NGC\,4850 & S0     &  3.66$\pm$0.69 & 1 & 1.00 &  4.81$\pm$2.41 & 1 & 1.00 & 11.54$\pm$0.1 & 15.22$\pm$0.1\\
160070 & NGC\,4854 & S0     &  8.51$\pm$1.05 & 1 & 1.00 & 18.28$\pm$2.70 & 1 & 1.00 & 11.40$\pm$0.1 & 14.98$\pm$0.1\\
160103 & NGC\,4926 & E      &  7.25$\pm$0.98 & 1 & 1.00 &  9.37$\pm$2.25 & 1 & 1.00 & 10.25$\pm$0.1 & 14.08$\pm$0.1\\
160130 & NGC\,4957 & E      & 20.03$\pm$0.84 & 1 & 1.00 & 14.52$\pm$2.70 & 1 & 1.00 & 10.44$\pm$0.1 & 14.07$\pm$0.1\\
160215 & NGC\,4860 & E      &  6.69$\pm$0.21 & 1 & 1.00 &  7.11$\pm$2.41 & 1 & 1.00 & 10.82$\pm$0.1 & 14.57$\pm$0.1\\
160217 & IC\,3957  & E      &  2.16$\pm$0.51 & 1 & 1.00 &  4.81$\pm$2.30 & 1 & 1.00 & 11.95$\pm$0.1 & 15.50$\pm$0.1\\
160218 & IC\,3959  & E      &  2.88$\pm$0.60 & 1 & 1.00 &  5.52$\pm$2.25 & 1 & 1.00 & 11.35$\pm$0.1 & 15.52$\pm$0.1\\
160220 & IC\,3963  & S0     & 12.85$\pm$3.17 & 1 & 1.00 &  5.91$\pm$2.41 & 1 & 1.00 & 11.92$\pm$0.1 & 15.69$\pm$0.1\\
160221 & NGC\,4864 & E      &  5.89$\pm$0.36 & 1 & 1.00 &  6.79$\pm$2.25 & 1 & 1.00 & 11.11$\pm$0.1 & 14.45$\pm$0.1\\
160222 & NGC\,4867 & E      &  4.43$\pm$0.81 & 1 & 1.00 &  3.82$\pm$2.25 & 1 & 1.00 & 11.68$\pm$0.1 & 15.45$\pm$0.1\\
160229 & NGC\,4873 & S0     &  9.95$\pm$0.80 & 1 & 1.00 &  6.64$\pm$2.41 & 1 & 1.00 & 11.66$\pm$0.1 & 15.41$\pm$0.1\\
160230 & NGC\,4872 & S0     &  2.57$\pm$0.27 & 1 & 1.00 &  3.10$\pm$2.35 & 1 & 1.00 & 11.66$\pm$0.1 & 15.26$\pm$0.1\\
160234 & NGC\,4876 & E      &  6.85$\pm$0.83 & 1 & 1.00 &  4.92$\pm$2.25 & 1 & 1.00 & 11.61$\pm$0.1 & 15.47$\pm$0.1\\
160237 & NGC\,4883 & S0     &  7.89$\pm$0.79 & 1 & 1.00 &  6.05$\pm$2.41 & 1 & 1.00 & 11.51$\pm$0.1 & 15.35$\pm$0.1\\
160238 & NGC\,4881 & E      &  5.96$\pm$0.60 & 1 & 1.00 &  9.59$\pm$2.25 & 1 & 1.00 & 10.89$\pm$0.1 & 14.61$\pm$0.1\\
160239 & NGC\,4882 & E      & 12.28$\pm$1.01 & 1 & 1.00 &  9.16$\pm$2.30 & 1 & 1.00 & 11.52$\pm$0.1 & 15.22$\pm$0.1\\
160242 & IC\,4011  & E      &  4.48$\pm$0.51 & 1 & 1.00 &  4.09$\pm$2.30 & 1 & 1.00 & 12.40$\pm$0.1 & 16.14$\pm$0.1\\
160244 & IC\,4012  & E      &  2.47$\pm$0.14 & 1 & 1.00 &  3.03$\pm$2.41 & 1 & 1.00 & 11.91$\pm$0.1 & 15.89$\pm$0.1\\
160246 & IC\,4021  & S0     &  2.13$\pm$0.44 & 1 & 1.00 &  3.48$\pm$2.41 & 1 & 1.00 & 11.82$\pm$0.1 & 15.92$\pm$0.1\\
160247 & NGC\,4894 & S0     & 21.69$\pm$4.56 & 1 & 1.00 &  4.59$\pm$2.41 & 1 & 1.00 & 12.34$\pm$0.1 & 16.04$\pm$0.1\\
160250 & IC\,4026  & S0     &  5.79$\pm$0.64 & 1 & 1.00 &  6.79$\pm$2.41 & 1 & 1.00 & 11.88$\pm$0.1 & 15.61$\pm$0.1\\
160253 & NGC\,4906 & E      &  5.43$\pm$0.45 & 1 & 1.00 &  7.98$\pm$2.25 & 1 & 1.00 & 11.54$\pm$0.1 & 15.30$\pm$0.1\\
160254 & IC\,4041  & S0     & 11.94$\pm$2.52 & 1 & 1.00 &  5.91$\pm$2.41 & 1 & 1.00 & 12.04$\pm$0.1 & 15.56$\pm$0.1\\
160256 & IC\,4045  & E/S0   &  5.35$\pm$0.66 & 1 & 1.00 &  4.81$\pm$2.25 & 1 & 1.00 & 11.25$\pm$0.1 & 14.99$\pm$0.1\\
160259 & IC\,4051  & E      & 15.60$\pm$0.72 & 1 & 1.00 & 11.53$\pm$2.25 & 1 & 1.00 & 10.78$\pm$0.1 & 14.40$\pm$0.1\\
\end{tabular}
\end{minipage}
\end{table*}

In Tab. 3, we list the properties of the 29 giant E and S0 galaxies
under study relevant to this analysis, as follows:
\newline
Col. 1: the galaxy denomination; \newline
Col. 2: alternate (NGC/IC) galaxy denomination; \newline
Col. 3: the morphological classification; \newline
Col. 4: the H-band effective radius $r_{e H}$ and its error; \newline
Col. 5: the classification according to the H-band profile decomposition
(see Col. 5 of Tab. 2); \newline
Col. 6: the H-band bulge-to-total luminosity ratio $B/T_H$ (see Col. 6
of Tab. 2); \newline
Col. 7: the B-band effective radius $r_{e B}$ and its error; \newline
Col. 8: the classification according to the B-band profile decomposition
(see Col. 5 of Tab. 2); \newline
Col. 9: the B-band bulge-to-total luminosity ratio $B/T_B$ (see Col. 6
of Tab. 2); \newline
Col. 10: the observed total H-band magnitude H and its error; \newline
Col. 11: the observed total B-band magnitude B and its error.

Hereafter no correction for Galactic extinction in direction either of
the Virgo cluster or of the Coma cluster will be applied to the B- and H-band
magnitudes, since this correction is negligible for our purposes.
No correction for internal extinction to the photometric parameters
of individual galaxies will be applied either.

\section{Results}

\subsection{Wavelength-dependence of the effective radius
in early-type dwarfs} 

Fig. 2 shows the distribution of the 18 VCC early-type dwarfs
listed in Tab. 1 in the plane defined by the decimal logarithm
of the ratio of $r_{e B}$ and $r_{e H}$ ($r_{e B}/r_{e H}$)
and by the total color index B$-$H.
Individual galaxies are represented by empty circles,
asterisks or filled circles if their surface brightness profile
follows either a de Vaucouleurs-law, a bulge$+$disk decomposition,
or an exponential-law (type 1, 2 or 3, respectively -- cf. Sect. 2).
In panels `a' and `b' objects are classified according to profile
decomposition either in the H-band or in the B-band (cf. Tab. 2),
respectively.
Fig. 2 shows that:
\begin{itemize}
\item
$r_{e B}/r_{e H}$ spans the broad range 0.7--2.2;
\item
7 early-type dwarfs with a relatively red color
(i.e. with $3.2 <$ B$-$H $< 4$), i.e. VCC\,608, 965, 1036, 1173,
1254, 1348 and 1491, have an average value of $r_{e B}/r_{e H}$
equal to 1 (with an rms dispersion of 0.21);
\item
10 early-type dwarfs and 1 E pec/S0 galaxy with a blue color
(i.e. with $2.5 <$ B$-$H $< 3.1$) have B-band effective radii
about 50\% longer than the H-band ones, on average
(with an rms dispersion of 0.44);
\item
when removing the three objects with a blue central excess (G01),
i.e. VCC\,951, 1499 and 1684, and the dE pec? galaxy VCC\,1078,
the remaining subsample of 7 blue early-type dwarfs
clusters around an average value of $r_{e B}/r_{e H} = 1.75$
(with an rms dispersion of 0.35).
\end{itemize}

Blue and red objects seem to be separated by a kind of ``gap'' in Fig. 2.
We are not able to trace the origin of this apparent gap back to
the selection effects of the G01 sample; instead we discuss the hypothesis
that this gap is an effect of data analysis.
\begin{figure}
 \vspace{400pt}
 \caption{Plot of the total color index B$-$H vs. the decimal logarithm
  of the ratio of the B-band effective radius ($r_{e B}$)
  and the H-band effective radius ($r_{e H}$), $r_{e B}/r_{e H}$,
  for the 18 VCC early-type dwarfs (see Tab. 2).
  Galaxies are identified according to either the H-band profile
  classification given in Col. 5 of Tab. 2 ($\bf a$) or the B-band profile
  classification given in Col. 10 of Tab. 2 ($\bf b$).
  In each panel, the 18 VCC early-type dwarfs are represented by
  empty circles, asterisks and filled circles if their surface brightness
  profile follows either a de Vaucouleurs-law, a bulge$+$disk decomposition,
  or an exponential-law (type 1, 2 or 3, respectively -- cf. Sect. 2).}
\end{figure}
First, we note that the classification according to surface brightness
profile decomposition does not change from the H-band to the B-band,
for the 7 red early-type dwarfs.
By contrast, 6 out of the 11 blue early-type dwarfs change
their classification from the near-IR to the optical case.
Among these 6 objects, VCC\,745, 1122, 1308 and 1386 may rise
the suspect of a missed/misidentified exponential-disk component
(cf. Sect. 2).
The values of $r_{e H}$ of these 4 galaxies should be underestimated
by 40--90\% in order to erase the difference between $r_{e H}$ and $r_{e B}$,
while the correction of $r_{e H}$ supported by inspection of
the H-band photometric profiles (G01, their Fig. 4) is 10\%.
The magnitudes of these corrections (if due) may shift the distribution
of the previous 4 galaxies toward smaller values of $r_{e B}/r_{e H}$
and redder B$-$H colors, filling the apparent gap in Fig. 2,
but do not weaken the trend seen there.
Hereafter we do not consider this gap as a significant feature.
\begin{figure}
 \vspace{400pt}
 \caption{B$-$H vs. $r_{e B}/r_{e H}$ for the 18 VCC early-type dwarfs,
  here classified according to the slope of their B$-$H ($\bf a$)
  or B$-$V($\bf b$) color profiles (cf. G01, their Fig. 4). Empty squares,
  empty triangles, filled squares and sun symbols identify objects
  with negative, null, positive and complex color profile slopes,
  respectively.}
\end{figure}

In order to investigate if the average increase of $r_{e B}/r_{e H}$
with B$-$H consistently reflects the average behaviour of the B$-$H
color gradient with the total B$-$H color, we reproduce the same plot
of Fig. 2 in Fig. 3a, but by adopting a classification based on the slope
of the B$-$H color profile.
B$-$H and B$-$V color profiles of the dwarfs under study are determined
and displayed by G01 (their Fig. 4).
From the full extension of these profiles, we derive average gradients
in B$-$H and in B$-$V, in order to verify if the signs of these two color
gradients are consistent.
In general they are, and, as expected, the strength of the B$-$V color
gradient is much lower.
Nonetheless, in Fig. 3b, we plot B$-$H vs. $r_{e B}/r_{e H}$ for our dwarfs,
but we adopt a classification based on the slope of the B$-$V color profile,
if available.

Fig. 3a shows that, for 6 out of the 7 blue early-type dwarfs
with average $r_{e B}/r_{e H} = 1.75$, B$-$H decreases at larger
galactocentric distances.
We estimate strengths of the B$-$H color gradient
$\Delta$(B$-$H)/$\Delta log~r$ equal to -0.92, -0.63, -0.60, -0.57,
-0.37 and -0.33 $\rm mag~dex^{-1}$ for VCC\,745, 1453, 1073, 1308,
786 and 1386, respectively.
\begin{figure}
  \vspace{400pt}
  \caption{Plot of the bulge-to-total luminosity ratio, either in the H-band
  ($B/T_H$) ({\bf a}) or in the B-band ($B/T_B$) ({\bf b}) vs.
  $r_{e B}/r_{e H}$ for the VCC early-type dwarf sample.
  Objects are classified according to their H-band surface brightness profile
  decomposition and their symbols are the same as in Fig. 2a.}
\end{figure}
By contrast, out of the 7 red early-type dwarfs, only 2 object
(VCC\,1036 and 1491) have a negative slope of the B$-$H color profile,
while 4 objects have no gradient in B$-$H and 1 object (VCC\,1254)
becomes redder toward its outer regions.
For VCC\,1036 and 1491, we estimate values of $\Delta$(B$-$H)/$\Delta log~r$
equal to -0.33 and -0.20 $\rm mag~dex^{-1}$, respectively,
significantly smaller, in absolute value, than the average strength
of $\Delta$(B$-$H)/$\Delta log~r$ of the previous 6 blue early-type dwarfs.
None of the 3 blue early-type dwarfs with a blue central excess (G01),
has a negative B$-$H color gradient, so that the strong blue central excess
seems a sufficient explanation to their low values of $r_{e B}/r_{e H}$
within the subsample of blue early-type dwarfs.
We conclude that stronger negative color gradients are necessarily
associated with larger values of $r_{e B}/r_{e H}$.
However, since there is no reason a priori why strong negative
color gradients should be associated with blue colors, we conclude that
the displacement of the average locus of the 7 blue early-type dwarfs
without a blue central excess from that of the 7 red dEs and dS0s in Fig. 2
is a robust and interesting behaviour, though is only a $\rm 2 \sigma$ effect.
\begin{figure}
  \vspace{400pt}
  \caption{Plot of $B/T_H$ ({\bf a}) and $B/T_B$ ({\bf b}) vs. B$-$H
  for the 18 sample early-type dwarfs.
  Galaxies are classified according to their H-band surface brightness
  profile decomposition and their symbols are the same as in Fig. 2a.}
\end{figure}

Whatever the classification in terms of profile decomposition is
(e.g. in the H-band), $r_{e B}/r_{e H}$ does not seem to depend
on the bulge-to-total luminosity ratio, either in the H-band
or in the B-band, as shown in Fig. 4a,b, respectively.
Interestingly, among the 5 blue objects which show an exponential-disk
component only in the B-band surface brightness profile, the highest value
of $B/T_B$ and the smaller value of $r_{e B}/r_{e H}$ belong to
the E pec/S0 galaxy VCC\,1499.
According to G01, the very blue continuum and strong Balmer absorption lines
of this object are typical of E$+$A galaxies which have experienced
an intense burst of star formation ended about 1--2 Gyrs ago
(Poggianti \& Barbaro 1996).
\begin{figure*}
  \vspace{200pt}
  \caption{B$-$H color--H-band magnitude relation of the 18 VCC early-type
  dwarfs (large filled squares) and of the 29 CGCG early-type giants
  of the Coma cluster listed in Tab. 3 (small empty squares).}
\end{figure*}

\begin{figure*}
  \vspace{200pt}
  \caption{$r_{e B}/r_{e H}$ vs. H-band magnitude for the 18 VCC early-type
  dwarfs (large filled squares) and for 29 early-type giants
  of the Coma cluster (small empty squares). The solid line and the two
  short-dashed lines represent the mean (1.13) and the $\pm 1 \sigma$ limits
  of the distribution of $r_{e B}/r_{e H}$ of the giants, respectively.}
\end{figure*}
As well as $r_{e B}/r_{e H}$, B$-$H does not seem to depend
on the bulge-to-total luminosity ratio, as shown in Fig. 5,
where we plot $B/T_H$ vs. B$-$H (panel `a') and $B/T_B$ vs. B$-$H
(panel `b'), the VCC galaxies being classified according to their H-band
surface brightness profile decomposition (cf. Fig. 2a).

Both $B/T_H$ and $B/T_B$ are less than 0.5 for those galaxies
with a confirmed exponential-disk component in both pass-bands,
so that, in these galaxies, the stellar populations associated
with the disk largely contribute to, if not dominate, the emission
in the near-IR and optical.
Finally, we note that $B/T_H$ and $B/T_B$ are to first order identical
both for the 7 red early-type dwarfs and for the previous 6 blue
early-type dwarfs.

\subsection{Early-type galaxies: dwarfs vs. giants}

In this section we investigate if the increase of $r_{e B}/r_{e H}$
with B$-$H previously found in the early-type dwarfs has an analogy
in giant E and S0 galaxies.
The findings may cast some light to the origin of the trend
illustrated in Fig.2.
As a comparison, we adopt the sample of 29 early-type giants
of the Coma cluster listed in Tab. 3.
Unfortunately, B$-$H or B$-$V color profiles of these giant galaxies
are not available to us, so that we can not study the behaviour
of total color, color gradient and $r_{e B}/r_{e H}$ in these systems.
\begin{figure*}
  \vspace{300pt}
  \caption{B$-$H vs. $r_{e B}/r_{e H}$ for the 18 VCC early-type
  dwarfs (large filled squares) and for 29 early-type giants
  of the Coma cluster (small empty squares). The solid line and the two
  short-dashed lines represent the mean (1.13) and the $\pm 1 \sigma$ limits
  of the distribution of $r_{e B}/r_{e H}$ of the giants, respectively.}
\end{figure*}

Given that the range of $\sim$ 600 in H-band luminosity (i.e. mass)
spanned by early-type giants and dwarfs implies differences in structural,
kinematical and photometric properties, first we display the distribution
of early-type dwarfs and giants in the planes $\rm M_H$--B$-$H (Fig. 6)
and $\rm M_H$--$r_{e B}/r_{e H}$ (Fig. 7).
Each sample of galaxies spans a range of $\sim$ 10 in H-band luminosity
(Fig. 6), but the early-type giants, plotted as small empty squares,
span a narrow range in color ($3.3 <$ B$-$H $< 4.2$), which overlaps
the range in B$-$H spanned by the red subsample of the early-type dwarfs,
hereafter plotted as large filled squares.
B$-$H colors bluer than 3.1 are found only in early-type dwarfs.
Conversely, the early-type giants span a broad range in $r_{e B}/r_{e H}$,
their $r_{e B}/r_{e H}$ ranging from $\sim 0.2$ to $\sim 2.2$,
with a mean value equal to 1.13 and an rms dispersion of 0.48 (cf. Fig. 7).
Such a broad range, due in part to the non-homogeneous source
of the photometric parameters of the giants (cf. Sect. 2), embraces
the distribution of the early-type dwarfs at the $\rm 2 \sigma$ level.
No trend of $r_{e B}/r_{e H}$ with H-band magnitude is found,
either overall or within each sample.

Finally, we reproduce the distribution of the early-type giants and dwarfs
in the $r_{e B}/r_{e H}$--B$-$H plane in Fig. 8, where the solid line
and the two short-dashed lines represent the mean (1.13)
and the $\pm 1 \sigma$ limits of the distribution
of $r_{e B}/r_{e H}$ of the giants, respectively.
Fig. 8 shows that the trend of increasing $r_{e B}/r_{e H}$ with bluer B$-$H,
found for dwarf systems, is not reproduced in the case of giant systems.
This result, if not due to the larger uncertainties affecting
the photometric quantities of the early-type giants under study,
points to the conclusion that the wavelength-dependence of the effective
radius of these stellar systems either has various origins or is due
to a cause which does not modify the total color in a consistent,
detectable manner.

The solid line representing the mean value of $r_{e B}/r_{e H}$
of the giants (1.13) bisects the distribution of the dwarfs:
6 out of the 7 red dwarfs have $r_{e B}/r_{e H}$ lower than 1.13
and, vice-versa, 8 out of the 11 blue dwarfs have $r_{e B}/r_{e H}$
larger than 1.13.
Moreover, giant E and S0 galaxies and blue early-type dwarfs show
maximum values of $r_{e B}/r_{e H} \sim 2.2$, despite the difference
in their colors.
Is therefore the trend claimed for the dwarfs a spurious result?
A definitive answer to this question may come only from an analogous study
based on larger statistics.
Here we note that a relation between total colors and sign and strength
of color gradients, as suggested for the early-type dwarfs under study
(cf. Sect. 3.1), is an intriguing result and its existence is consistent with
some scenarios of dwarf galaxy formation and evolution, as discussed next.

\section{Three different interpretations}

The increase of $r_{e B}/r_{e H}$ in blue early-type dwarfs (Fig. 2)
reflects the existence of a (strong) negative gradient
in B$-$H and in B$-$V along the radial coordinate (Fig. 3).
The data show that this is a necessary but not sufficient condition however.
This trend is not found in giant E and S0 galaxies,
which span a broad range in $r_{e B}/r_{e H}$, though their B$-$H color
is almost as red as in red early-type dwarfs.
Radial color gradients observed in individual giant elliptical and lenticular
galaxies are commonly interpreted as a product either of age/metallicity
gradients of the stellar populations along the radial coordinate
or of dust attenuation (cf. Sect. 1).
These interpretations do not imply a correlation between $r_{e B}/r_{e H}$
and B$-$H.

In the following sections, these three theoretical interpretations
will be extended to our results for early-type dwarfs
and discussed individually, in relation with some scenarios
of early-type dwarf galaxy formation and evolution
which may justify the trend seen in Fig. 2.
The lack of data does not allow us to achieve a direct proof
of the validity of any of these three interpretations.
Moreover, the absence of a one-to-one correlation between
$r_{e B}/r_{e H}$ and the strength of the gradient in B$-$H,
if not due to errors in data analysis, may indicate that the interpretation
of Fig. 2 is complex.

\subsection{Age gradients}

The integrated broad-band colors of E and S0 galaxies become progressively
bluer toward fainter magnitudes (Faber 1973; see however Scodeggio 2001).
This correlation, known as the color--magnitude relation,
is universal and very tight in the optical for ellipticals and lenticulars
in clusters at z=0 (Bower, Lucey \& Ellis 1992a,b).
Relying on the commonly accepted interpretation of the color--magnitude
relation (Kodama \& Arimoto 1997), we conclude that, in blue early-type
dwarfs, either the average metallicity is lower than in red early-type dwarfs
of the same H-band luminosity or star formation is still going on.
Here we make the hypothesis that the blue colors of the VCC sample dwarfs
are due to the presence of a young stellar population.
This hypothesis is supported by spectroscopy (G01) for some individual
sample galaxies and is consistent with the results of Terlevich et al. (1999).

Fig. 5b shows that most of the B-band emission of the dwarfs
is contributed by the stellar populations distributed within
an exponential-disk component.
This disk-component, if present, is also responsible for more than 50\%
of the total H-band luminosity (Fig. 5a).
If we make the assumption that early-type dwarfs are rotationally flattened
(cf. Sect. 1), the wavelength-dependence of their effective radius,
and the relation between total color and sign and strength of
the color gradients may be understood as effects of disk evolution.
In this scenario, an increase of $r_{e B}/r_{e H}$ happens in objects
with on-going star formation activity, when the star formation, viscous
and gaseous infall timescales are comparable (cf. Sect. 1).
By contrast, the effective radius is not expected to change with wavelength
in objects where either the bulk of the disk (and of the system) is assembled
before star formation and viscosity act in a significant manner
or star formation has ceased.
Therefore, the objects with $r_{e B}/r_{e H} \sim 1$ are expected to have
red colors, as we find.

If red and blue early-type dwarfs belong to the same population (cf. Sect. 1),
the former should be quiescent since more than 1--2 Gyrs, in order to justify
the difference in colors between e.g. VCC\,1491 and 1499 (G01).
This quiescence may be due either to a relatively shorter star formation
timescale or to an early interruption of the gaseous infall, due to phenomena
like galactic winds, tidal stripping or ram-pressure by the intergalactic
medium (IGM).
There are no detections of HI gas for the VCC early-type dwarfs under study,
but only upper limits to the observed HI flux for VCC\,1254 and 1499
(Huchtmeier and Richter 1986), VCC\,951 and 1491 (van Driel et al. 2000).
Interestingly, increasing upper limits are associated with bluer objects.
No trend between color and kinematic distribution is found
for the 15 objects with measured heliocentric velocities.
Beside this, Fig. 1 shows that all the early-type dwarfs
of our sample but VCC\,608 lay in dense regions of the Virgo cluster.
These findings disfavour a key-role of IGM ram-pressure.
On the other hand, we note that the fraction of dE and dS0 galaxies
with at least one giant galaxy within a projected distance of 15 arcmin
(equivalent to $\rm \sim 75~kpc$ at the assumed distance of Virgo) drops
from 57\% to 18\%, when the red and blue subsamples are considered
separately, respectively.
This seems to favour tidal stripping as a mechanism of gas removal
in the red VCC early-type dwarfs.
Conversely, gas infall may sometimes trigger bursts
of star formation in the nuclear regions of a galaxy (e.g. VCC\,1499).
The associated increase in blue luminosity in this central region
may eventually erase differences between optical and near-IR effective radii.

\subsection{Metallicity gradients}

The distribution in Fig. 2 may be explained by different strengths
of a metallicity gradient along the radial coordinate in individual
stellar systems with an assumed single-age stellar population
(e.g. Bruzual \& Charlot 1993; Worthey 1994).
Such an interpretation has been invoked by Peletier, Valentijn \& Jameson
(1990) and more recently by Tamura et al. (2000) and Saglia et al. (2000)
for (giant) elliptical galaxies.
For the sake of correctness, we say that metallicity gradients
are the currently most widely accepted interpretation of the existence
of color gradients in such systems.
On the other hand, we note that no correlation between color gradients
and e.g. $\rm Mg_2$ gradients has been found so far, though only a small
number of elliptical galaxies has been studied both for color gradient
and $\rm Mg_2$ gradient (Peletier 1989).
Reasons are discussed by Kobayashi \& Arimoto (1999).

Unfortunately, there is no information about the content
and radial distribution of metallicity of the early-type dwarfs
under study, but we may gain some understanding from the giants.
In fact, 14 out of the 29 giant E and S0 galaxies have radial profiles
of line indices like $\rm Mg_2$ thanks to the spatially resolved
spectroscopy of Mehlert et al. (2000).
Twelve of these galaxies have negative $\rm Mg_2$ line-strength gradients:
their colors and values of $r_{e B}/r_{e H}$ are consistent with those
of the rest of the giants.
If taken as a face value, the mean $r_{e B}/r_{e H} = 1.13$
obtained for the early-type giants indicates that metallicity gradients
may increase $r_{e B}$ by 10--15\% with respect to $r_{e H}$.
May the larger (up to 75\% on average) effect on $r_{e B}/r_{e H}$
seen for 7 blue early-type dwarfs be due to stronger metallicity gradients
in these galaxies?
A basic assumption to producing color and line-strength gradients
in elliptical galaxies is that galactic winds blow later in the inner part
of a galaxy because of a deeper potential well, defined mainly by dark matter
(Martinelli, Matteucci \& Colafrancesco 1998). 
However, efforts of fully reproducing observed color and line-strength
gradients failed so far (Tantalo et al. 1998), so that some important physics
seems to be missing.
This conclusion is supported by the fact that our blue and red early-type
dwarfs have comparable total masses (Fig. 6) and, reasonably,
central potential wells.

\subsection{Dust attenuation}

Giant elliptical and lenticular galaxies have been known to contain dust
since a long time, thanks to catalogues of far-IR emission 
(Jura 1986; Bally \& Thronson 1989; Knapp et al. 1989; Roberts et al. 1991),
extensive visual surveys for dust obscuration (e.g. Hawarden et al. 1981;
Ebneter \& Balick 1985; Sparks et al. 1985; Kim 1989) and the discovery
of atomic and molecular line emission (e.g. Sage \& Wrobel 1989;
Thronson, Bally \& Hacking 1989; Wiklind \& Henkel 1989; Gordon 1990;
Roberts et al. 1991; Lees et al. 1991).
Michard (1998) noted that dust patterns are more common and important
in disky ellipticals than in boxy ones.
Recently atomic line emission has been detected in early-type dwarfs
(e.g. Young \& Lo 1997), as well as dust signatures (Elmegreen et al. 2000).
The latter discovery is particularly relevant to this study
since it involves a dE,N galaxy of the Virgo cluster (VCC\,882).
It supports the hypothesis that color gradients and distribution
in $r_{e B}/r_{e H}$ of our early-type dwarfs are due to
differences in optical depth (i.e. dust column density) and/or dust geometry
among individual objects (Witt, Thronson \& Capuano 1992).
As Witt, Thronson \& Capuano point out, the geometry of the dust distribution
with respect to stars has an important role, due to the importance exercited
on scattering and on the determination of the relative fraction of stars
which are lightly obscured.
For one plausible geometry, these authors find that the maximum reddening
occurs at intermediate optical depths, while both very small and very large
amounts of dust produce almost neutral broad-band colors.
Transfer of radiation through a dusty medium with an optical depth $\tau_V$
of the order of 1 is able to reproduce some color and surface brightness
variations in giant elliptical galaxies that are usually attributed
to age/metallicity gradients of the stellar populations.

According to the results of Witt, Thronson \& Capuano, no relationship
of the dust-induced increase of $r_{e B}/r_{e H}$ with B$-$H must be expected
a priori, so that the trend in Fig. 2 might be due simply to chance.
On the other hand, since dust production is a result of star formation,
relatively larger amounts of dust may be associated with galaxies
of intrinsic blue colors.
In addition, a necessary condition for retainsion of metals
in dwarf stellar systems (Legrand et al. 2001), despite their higher
sensitivity to galactic winds than giant ones (Arimoto \& Yoshii 1987),
seems to be the presence of a diffuse gaseous halo
(D'Ercole \& Brighenti 1999; Silich \& Tenorio-Tagle 2001).
The hint that the maximum HI content allowed by observations increases
in bluer objects (cf. Sect. 4.1) may suggest that a greater part
of the metals (and of the dust) was retained by the relatively HI-richer
and bluer early-type dwarfs of our sample.
This may explain the trend of $r_{e B}/r_{e H}$ with B$-$H in Fig. 2.

With the caveat of large observational uncertainties, Fig. 8  suggests
a stronger effect of dust attenuation in the blue early-type dwarfs
than in most of the giant E and S0 galaxies.
This may sound at odd with the metal-richness of early-type giants
with respect to dE and dS0 galaxies (cf. Yoshii \& Arimoto 1987),
commonly interpreted with the galactic-wind model of Arimoto \& Yoshii (1987).
The interpretation of Fig. 8 in terms of dust content and distribution
is complex enough to deserve a quantitative investigation of the effect
of dust attenuation on the photometric properties on both giant
and dwarf E and S0 galaxies in a future paper.

\section{Conclusions}

Thanks to the surface photometry of 18 Virgo cluster dwarf elliptical (dE)
and dwarf lenticular (dS0) galaxies, made by Gavazzi et al. (2001)
in the H-band ($\rm 1.65~\mu m$) and in the B-band ($\rm 0.44~\mu m$),
we are able to study the wavelength-dependence of their effective radii.

We find that the ratio of the effective radii of these stellar systems
in these two spectral pass-bands, $r_{e B}/r_{e H}$, ranges between
0.7 and 2.2.
$r_{e B}/r_{e H}$ does not depend on the bulge-to-total light ratio
or total mass but on the total color index B$-$H.
In fact, dwarf ellipticals and lenticulars with a red total color index B$-$H
(i.e. with $3.2 <$ B$-$H $< 4$) have equal effective radii
in the B and H pass-bands.
By contrast, blue (i.e. with $2.5 <$ B$-$H $< 3.1$) dEs and dS0s have
B-band effective radii about 50\% longer than the H-band ones, on average.
Consistently, strong negative color gradients are present
only in the B$-$H color profiles of the blue dwarfs of our sample.

The trend of increasing $r_{e B}/r_{e H}$ with bluer B$-$H is not confirmed
by a reference sample of 29 well-studied Coma cluster giant E and S0
galaxies, which span a narrow range in color ($3.3 <$ B$-$H $< 4.2$).
For these early-type giants,  $r_{e B}/r_{e H}$ spans a broad range
(0.2--2.2) and has a mean value of $\sim 1.1$.
The origin of the wavelength-dependence of the effective radii
in these stellar systems (e.g.: age/metallicity gradient along
the radial coordinate and dust attenuation) does not imply a consistent
trend of $r_{e B}/r_{e H}$ with B$-$H.

Assuming each of these three distinct interpretations of the orgin
of color gradients, we discuss the origin of the association
of strong negative color gradients with blue colors found
in the early-type dwarfs under study, in relation with current scenarios
of formation and evolution of dE and dS0 galaxies.

\section*{Acknowledgments}
The author is indebted to G. Gavazzi for providing him
with part of the data used in this study. \newline
He is also indebted to the anonymous referee for his very stimulating
comments.


\bsp

\label{lastpage}

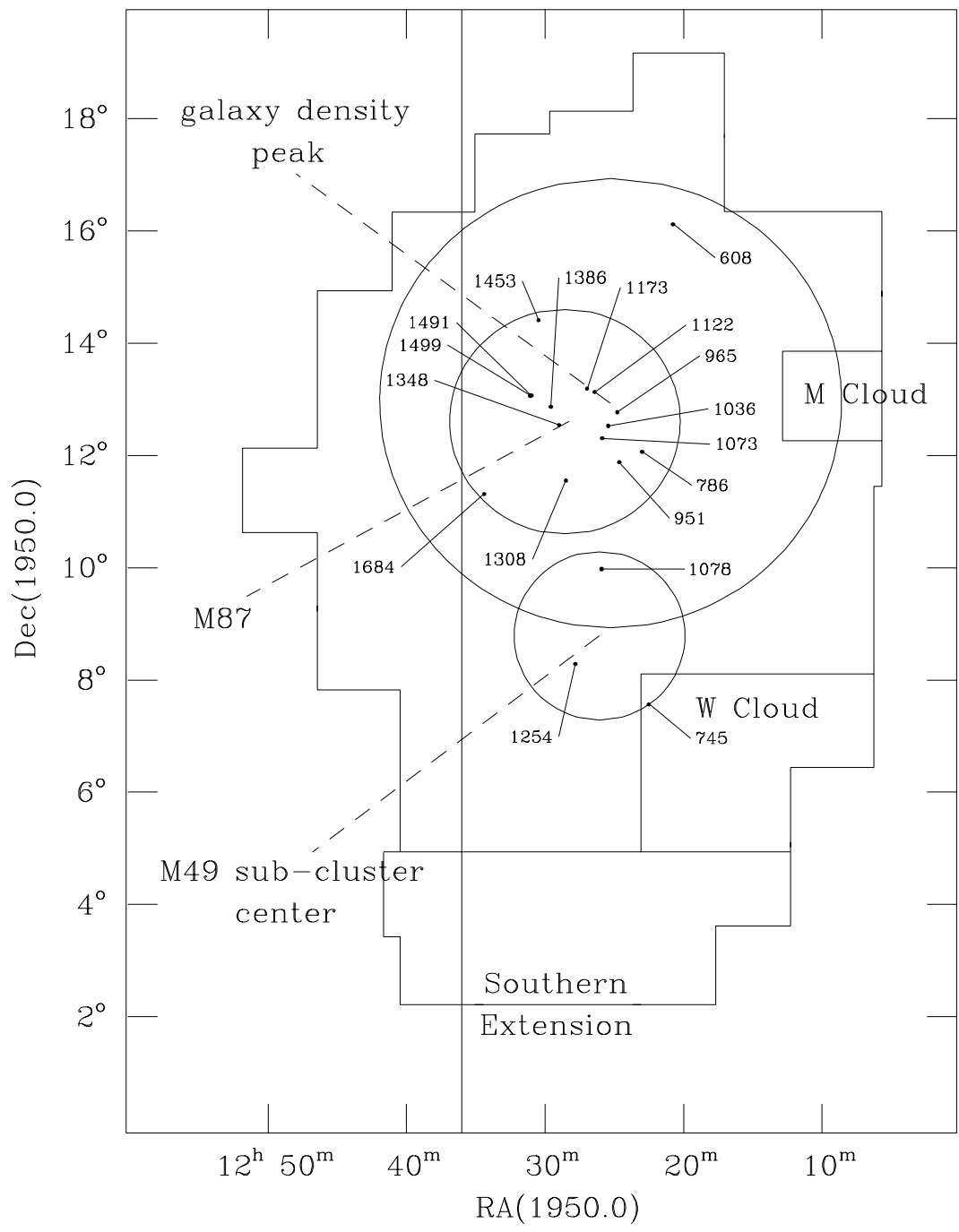
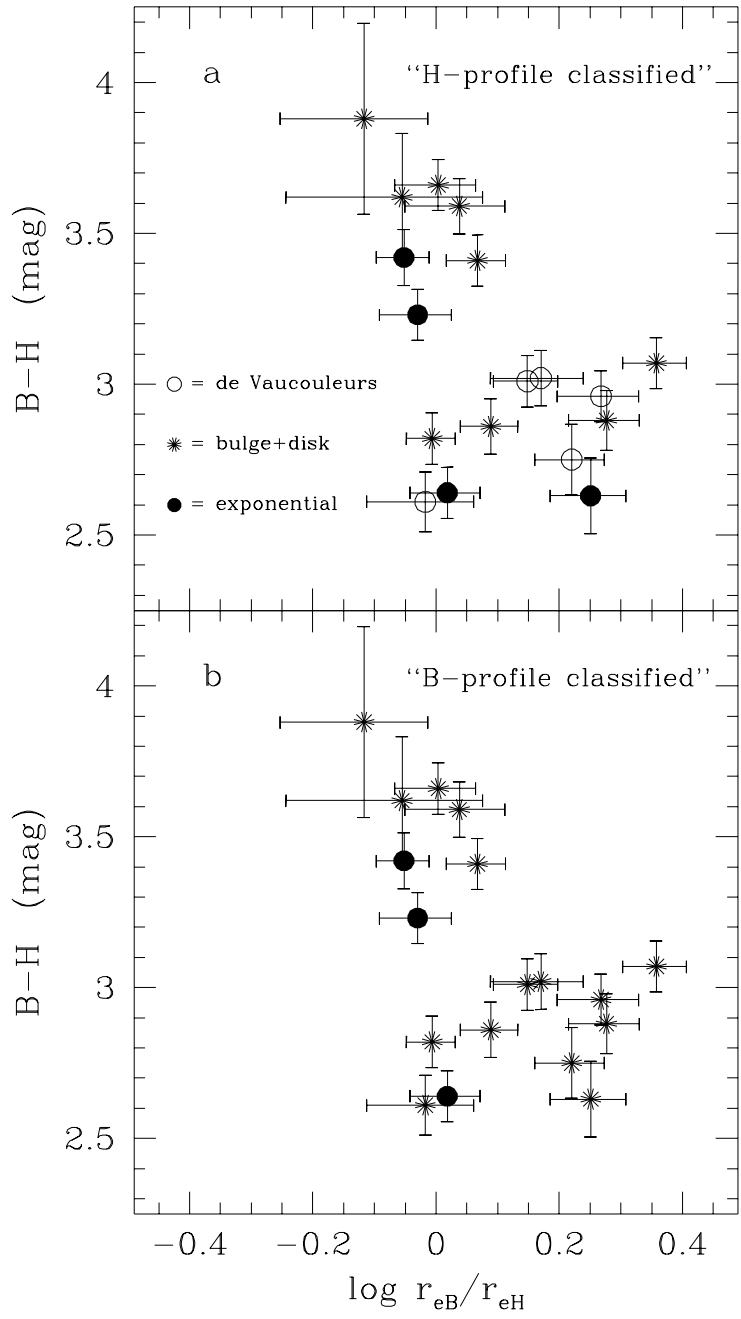
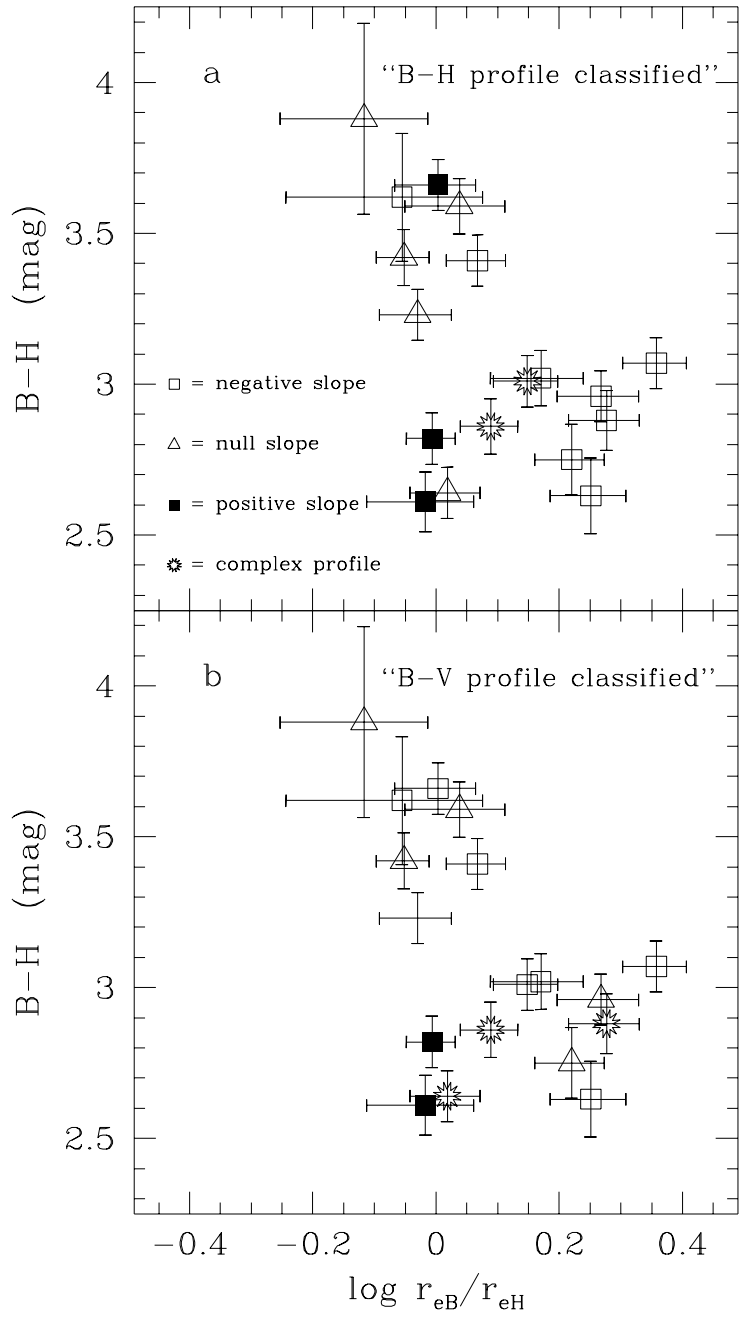
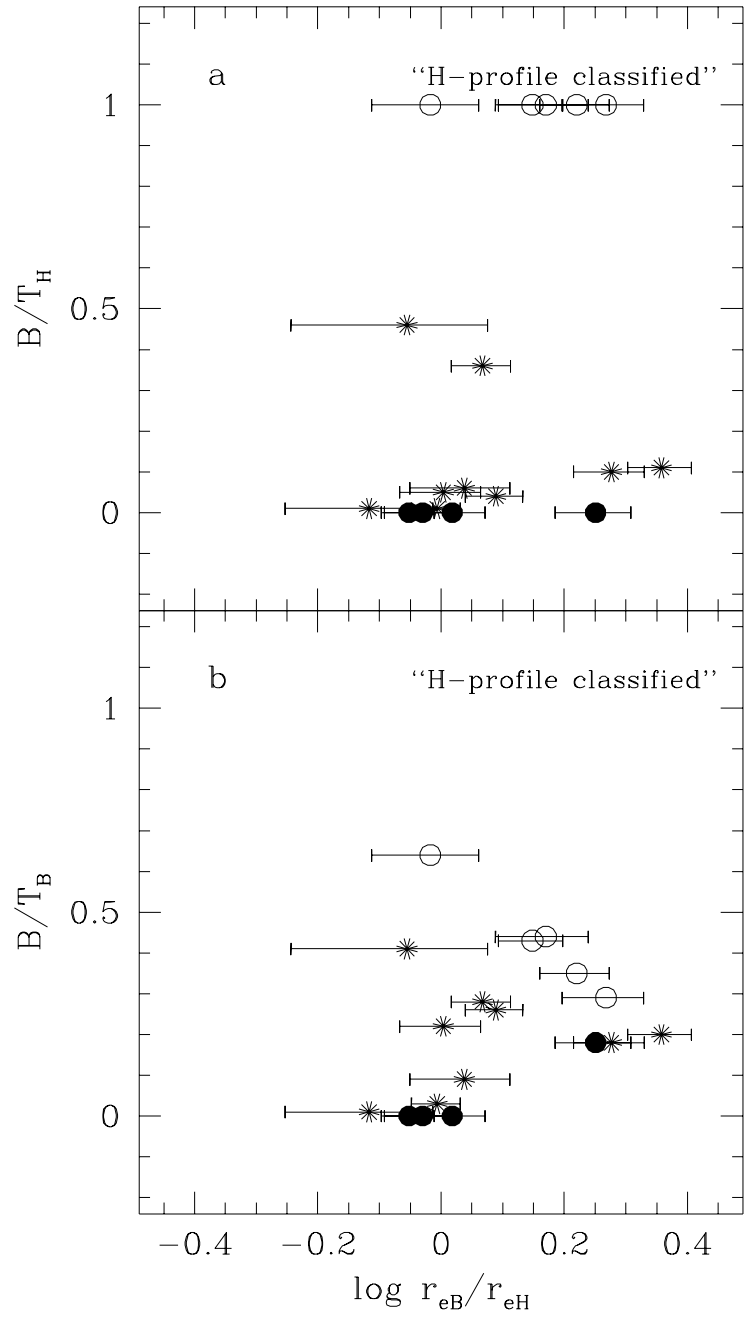
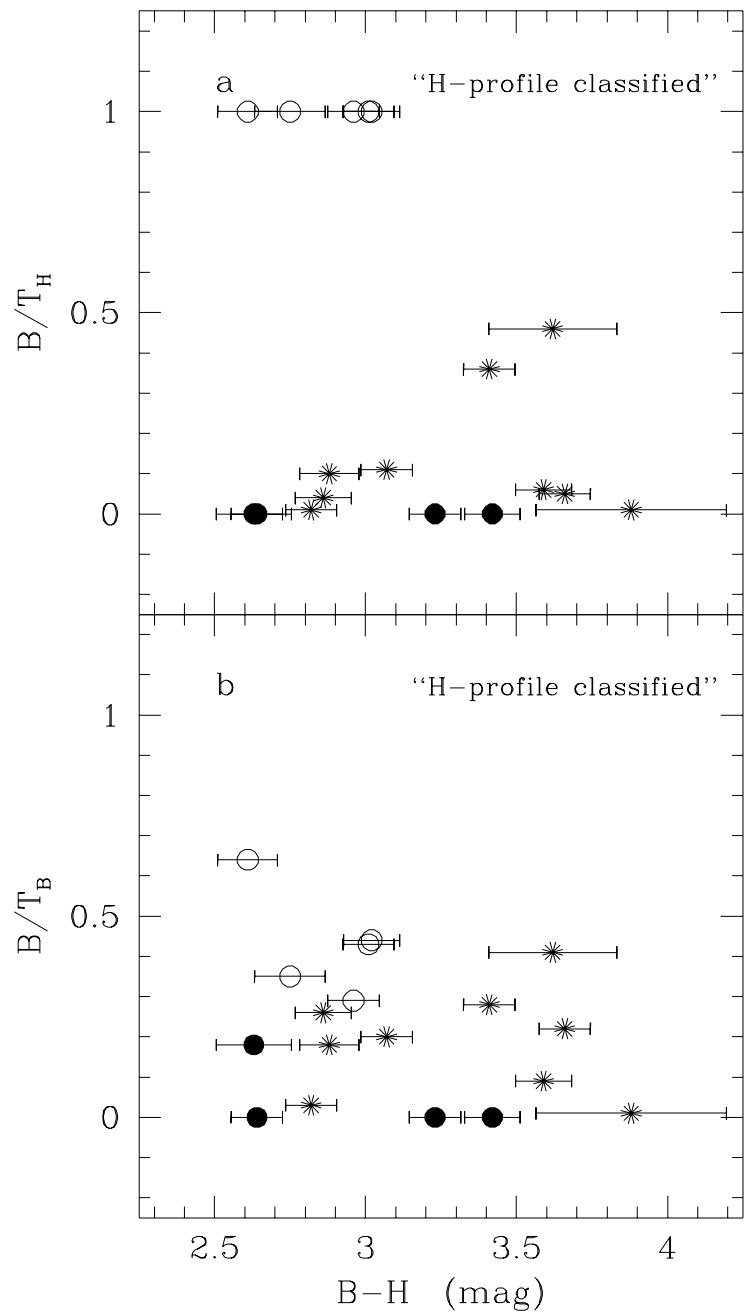
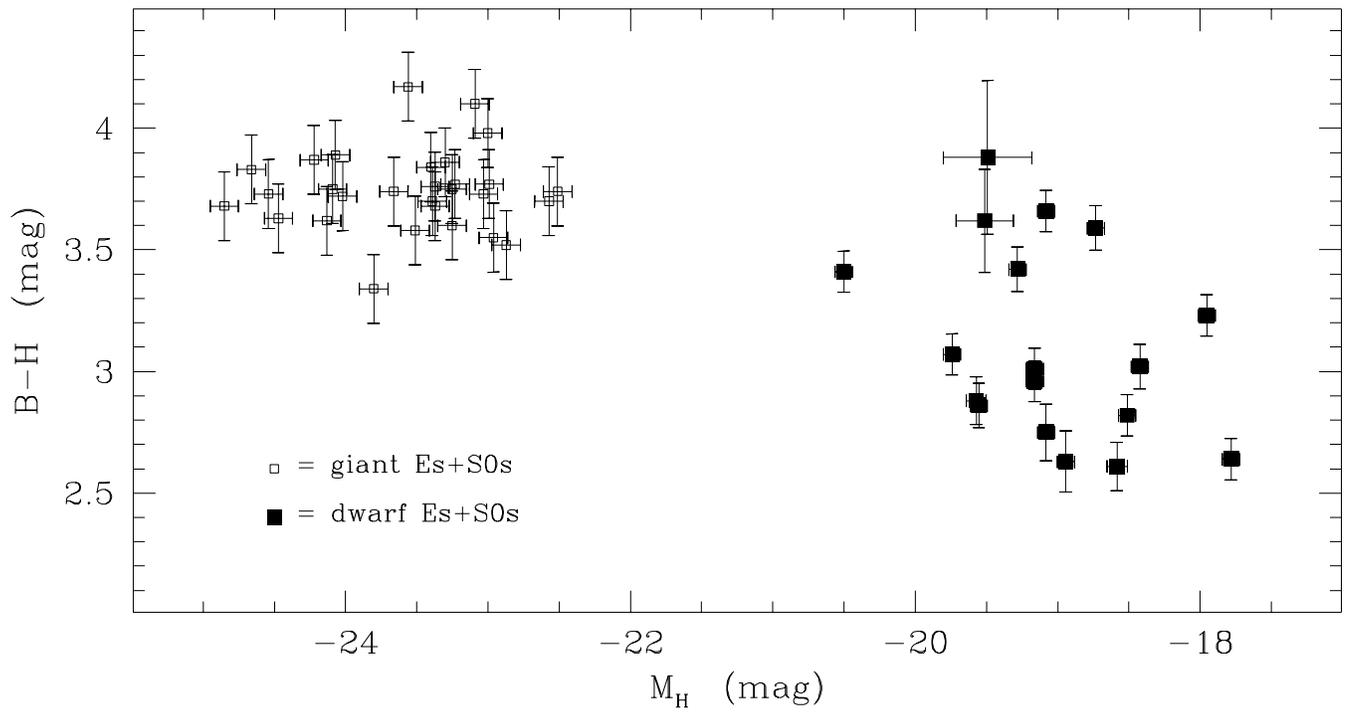
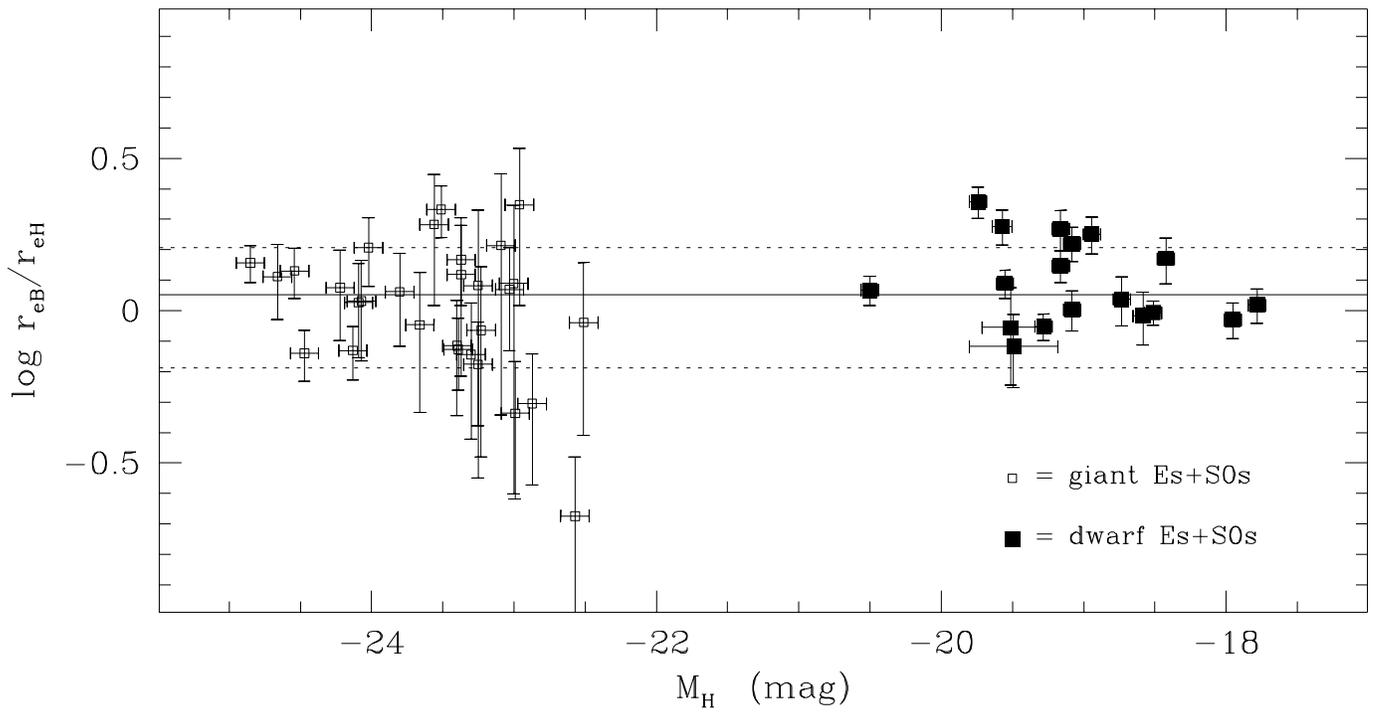
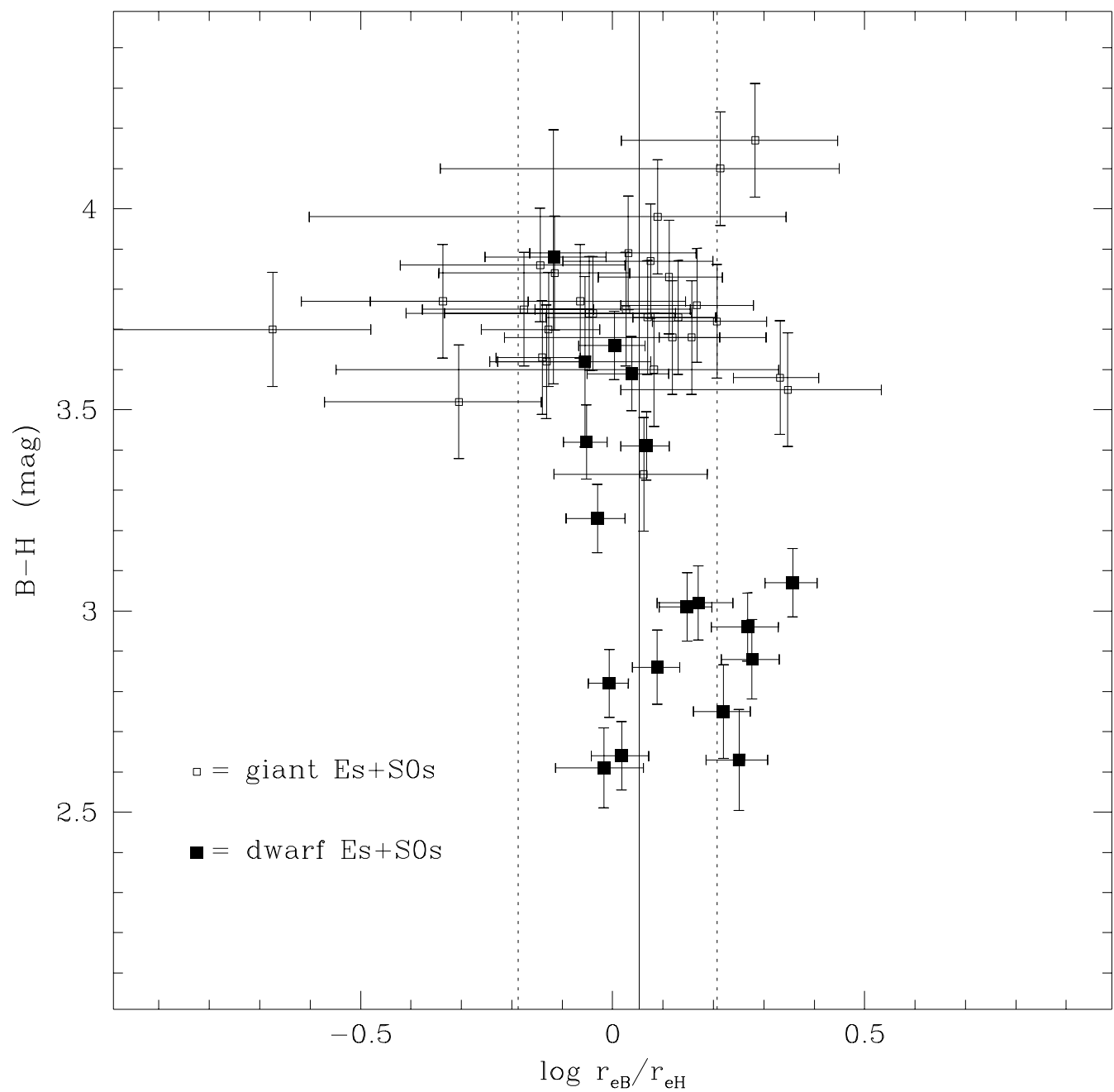
\end{document}